\documentclass[journal=jacsat,manuscript=article,layout=twocolumn]{achemso}
\PassOptionsToPackage{dvipsnames}{xcolor}
\usepackage{chemformula} % Formula subscripts using \ch{}
\usepackage[T1]{fontenc} % Use modern font encodings
\usepackage{booktabs}
\usepackage{subcaption}
\usepackage{xr}
\usepackage{lineno}
\let\oldmaketitle\maketitle
\let\maketitle\relax

\makeatletter
\newcommand*{\addFileDependency}[1]{% argument=file name and extension
\typeout{(#1)}% latexmk will find this if $recorder=0
\@addtofilelist{#1}
\IfFileExists{#1}{}{\typeout{No file #1.}}
}\makeatother

\newcommand*{\myexternaldocument}[1]{%
\externaldocument{#1}%
\addFileDependency{#1.tex}%
\addFileDependency{#1.aux}%
}

\myexternaldocument{si}

\author{Megan C. Davis}
\affiliation{Theoretical Division, Los Alamos National Laboratory, Los Alamos, NM, 87545, United States}
\alsoaffiliation{Center for Non-Linear Studies, Los Alamos National Laboratory, Los Alamos, NM, 87545, United States}
\email{megand@lanl.gov}
\author{Wilton J. M. Kort-Kamp}
\affiliation{Theoretical Division, Los Alamos National Laboratory, Los Alamos, NM, 87545, United States}
\author{Ivana Matanovic}
\affiliation{Theoretical Division, Los Alamos National Laboratory, Los Alamos, NM, 87545, United States}
\author{Piotr Zelenay}
\affiliation{Material Physics and Applications Division, Los Alamos National Laboratory, Los Alamos, NM 87545, United States}
\author{Edward F. Holby}
\affiliation{Theoretical Division, Los Alamos National Laboratory, Los Alamos, NM, 87545, United States}
\email{holby@lanl.gov}

\title[An \textsf{achemso} demo]
  {Design of Amine-Functionalized Materials for Direct Air Capture Using Integrated High-Throughput Calculations and Machine Learning}

\abbreviations{IR,NMR,UV}

\begin{document}

% \begin{tocentry}
% 	% \begin{figure}
% 	%   \centering
% 	\begin{center}
% 		\includegraphics[width=\textwidth]{toc/toc_figure_finished.png}
% 	\end{center}
% \end{tocentry}

\twocolumn[
	\begin{@twocolumnfalse}
		\oldmaketitle
		\begin{abstract}
			Direct air capture (DAC) of carbon dioxide is a critical technology for mitigating climate change, but current materials face limitations in efficiency and scalability.
We discover novel DAC materials using a combined machine learning (ML) and high-throughput atomistic modeling approach.
Our ML model accurately predicts high-quality, density functional theory-computed CO$_{2}$ binding enthalpies for a wide range of nitrogen-bearing moieties.
Leveraging this model, we rapidly screen over 1.6 million binding sites from a comprehensive database of theoretically feasible molecules to identify materials with superior CO$_{2}$ binding properties.
Additionally, we assess the synthesizability and experimental feasibility of these structures using established ML metrics, discovering nearly 2,500 novel materials suitable for integration into DAC devices.
Altogether, our high-fidelity database and ML framework represent a significant advancement in the rational development of scalable, cost-effective carbon dioxide capture technologies, offering a promising pathway to meet key targets in the global initiative to combat climate change.

%%% Local Variables:
%%% mode: latex
%%% TeX-master: "../main"
%%% End:

		\end{abstract}
	\end{@twocolumnfalse}
]
Reducing CO$_{2}$ concentration in the atmosphere is necessary to meet the international climate goals of the Paris Agreement.\cite{ParisAgree}
However, current approaches suffer from materials and engineering issues that limit their scalability and economics.\cite{Keith2018,NASEMed2019,Rubin2015}
Direct air capture of CO$_{2}$ using solid sorbents is a promising way to reduce atmospheric concentrations.\cite{Patel_2017,Shi2020,Samanta2011,Choi2009}
Carbon dioxide captured in this manner could be made chemically useful via electrochemical CO$_{2}$ reduction,\cite{Aresta2010,Kibria2019,Zhu2021}~potentially in an integrated system.\cite{Gutierrez-Sanchez2022}
Through this process, CO$_{2}$ may be reduced to value-added chemicals, including C1 products such as carbon monoxide and methane as well as C2 products such as ethylene.\cite{Nitopi2019}
This approach could make closing the carbon loop through CO$_{2}$ capture and utilization economically feasible.\cite{Zhong2020,Chen2020,Kas2020,Jiang2018,Kungas2020}

Amine-based solid sorbents are a promising class of materials for achieving efficient direct air capture of carbon dioxide.\cite{Yu2016,Nguyen2023,Said2020,Hack_2022}
These sorbents may be prepared via physical impregnation of porous materials, chemical grafting of a support surface (such as functionalization of carbonaceous nanofibers~\cite{Irani_2017}), or in-situ polymerization of an inorganic support.\cite{Shi2020}
Chemisorption of CO$_{2}$ by these materials can result in significantly improved CO$_{2}$ capacity compared to conventional MOF or zeolite-based physisorption.\cite{Shi2020,Mukherjee2019,Huck2014,Sanz-Perez2016}
Importantly, amines offer a diverse chemical space that could be explored for optimization of key properties such as CO$_{2}$ binding strength and thermo-oxidative stability.\cite{Heldebrant2017,Sakwa-Novak2015,Feric2021,Choi2016,Choi2021,Nezam2021,Vu2021,Lin2013,Singh2020}
Some examples include the incorporation of electron-withdrawing groups to lower amine basicity combined with high steric hindrance to reduce heat of CO$_{2}$ adsorption and improve thermo-oxidative stability.\cite{Min2018,Lashaki2019}
However, current state-of-the-art amine-based solid sorbents typically only reach adsorption capacities of 2 mmol CO$_{2}$/g.
This value generally drops significantly after only a few capture/regeneration cycles due to degradation of the material.\cite{Shi2020}
CO$_{2}$ capacity and stability have been identified as the key factors for making these materials an economically feasible solution for direct air capture applications, which requires achieving < \$100/tCO$_{2}$ capture, the target of the U.S. Department of Energy's Carbon Negative Shot.\cite{NASEMed2019,CarbonNegShot}

Discovery of new amine-functionalized sorbents fulfilling these requirements can be greatly expedited by leveraging computational modeling.
Density functional theory (DFT) has been effectively used to model CO$_{2}$ sorption characteristics of amine molecules.\cite{Heldebrant2017,Lin2013,Chen2014,Wang2014,Gupta2019,Ostwal2011}
For instance, DFT calculations of binding energetics have been valuable for understanding CO$_2$ binding mechanisms for amine-based direct air capture materials\cite{Mebane2013,Mebane2013a,Li_2016} and DFT properties have shown good agreement with measured CO$_{2}$ binding properties\cite{Li_2016}.

Although DFT is a powerful tool for computational modeling, its computational expense is prohibitive for efficiently exploring the large chemical space of potential active sites for amine-based solid sorbents.
Namely, machine learning (ML) for chemical applications promises to significantly accelerate prediction of molecular properties and help explore this design space.\cite{Shilpa2023,Mater2019}
ML trained on DFT has been shown to be capable of approximating DFT values to high accuracy,\cite{Faber2017} and surrogate computational models have been used to explore the design spaces for solid sorbents and polymers.\cite{Rajendran2023,Wilbraham2018}

Furthermore, surrogate ML models allow for screening of millions of candidate chemistries from computational databases such as QM9 and GDB-17,\cite{Ruddigkeit2012,Ramakrishnan2014} thereby greatly accelerating the discovery of new molecules for targeted applications.
In particular, descriptor-based ML models show promise for materials discovery in applied energy research.\cite{De_Vos2024,Xie2023,Sun2023,Mohan2024}
Such models offer not just increased speed of prediction but can be used with explanatory methods to understand the model's decision-making, thus leading to theoretical insight and aiding in rational design via identified chemical properties.\cite{Wellawatte2023,Xin2023,SHAP,Lundberg2020}

Here, we develop descriptor-based ML surrogate models for materials discovery for the vital emerging technology of direct air capture.
We use an automated high-throughput DFT workflow that computes tens of thousands of CO$_{2}$ binding enthalpies of nitrogen-bearing molecules that may be embedded as active sites in carbonaceous polymers in order to optimize their CO$_{2}$ capacity.
With ML, we are able to quickly explore diverse chemistries from databases with millions of theoretically possible molecules.
Uniquely, we train our models on our high-level \emph{ab initio} DFT database, and are thus able to perform high-throughput screening for DFT-derived properties (e.g., binding enthalpy) specific to direct air capture materials.
By utilizing databases which cover unique and under-explored regions of chemical space, we have discovered truly novel materials for direct air capture of carbon dioxide.
Unlike previous works, we also use synthesizability metrics to identify experimentally promising structures from this exotic chemical space, increasing the likelihood that materials with optimized performance will be realizable in the laboratory.
Lastly, we use explanatory methods in tandem with our descriptor based approach to gain quantitative insight into the underlying chemical factors that contribute to the complex phenomenon of CO$_{2}$ adsorption.
Our work promises to accelerate the rational design of direct air capture materials by providing key guidance for their optimization.
Such tailored functional materials enabled by this work will be highly valuable tools for actively combating anthropogenic climate change and provide a valuable template for future functional materials discovery.

%%% Local Variables:
%%% mode: latex
%%% TeX-master: "../main"
%%% End:

\section{Results}
\subsection{High-Throughput Workflow and ML Model Performance}
\begin{figure*}[ht]
	\centering
	\includegraphics[width=\textwidth]{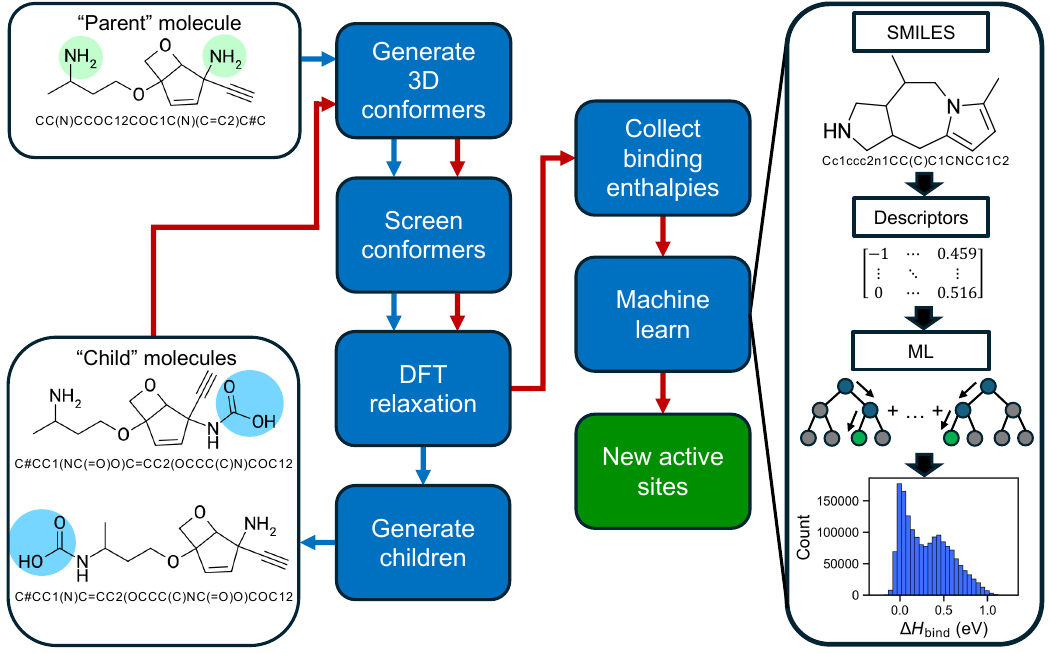}
	\caption{\label{fig:workflow} \small
		Workflow schematic demonstrating our automated high-throughput methodology.
		Following the blue arrows, conformers are generated using RDKit,\cite{RDKit} followed by relaxation with GFN2-xTB\cite{Bannwarth2019}.
		The minimum energy structures from this pre-relaxation step are passed to Gaussian 16 for DFT optimization\cite{g16} at the B3LYP/6-311+G(d,p) level of theory.
		We then use RDKit to generate structures for each of the ``children'' molecules based on the DFT-level geometry of the ``parent'' (active sites for the ``parent'' are highlighted in green, while bound CO$_{2}$ is highlighted in blue for the ``children'').
		The same steps outlined above are then repeated (red arrows).
		CO$_{2}$ binding enthalpies are then autonomously collected and stored using the pyiron framework.\cite{Janssen2019}
		Using scikit-learn~\cite{scikit-learn}, ML models are trained to predict DFT CO$_{2}$ binding enthalpies for a given molecule from an input vector of descriptors computed with the Mordred package.\cite{Moriwaki2018}}
\end{figure*}

Figure~\ref{fig:workflow} illustrates the high-throughput workflow used to generate a high-fidelity DFT database of CO$_{2}$ binding enthalpies for training our ML models.
We assume a carbamic acid formation mechanism to compute binding enthalpies for 15,336 nitrogen-bearing molecules from the GDB-17\cite{Ruddigkeit2012} and NIST\cite{Huber} databases, which correlated linearly (R$^{2}$ = 0.99) with a related ammonium carbamate mechanism (see Methods and Figure~\ref{si-fig:dma_trend}).
This dataset is used to train surrogate ML models by employing 2D descriptors from the Mordred package\cite{Moriwaki2018} as input features and CO$_{2}$ binding enthalpies as the target property for optimization.
2D descriptors are computed from string-based SMILES\cite{Weininger1988} representations of the molecules.
Our models performed best using differential descriptors, which are the difference between the carbamic acid (``child'') molecule and the original (``parent'') molecule (see Methods).
Lastly, our best performing trained model is used to predict binding enthalpies for 1,650,601 binding sites across 992,959 ``parent'' molecules from the GDB-17 database.
We also use SAscore\cite{Ertl_2009} and GDBscore\cite{Thakkar_2021} metrics to estimate synthesizability for each molecule.

\begin{figure}[ht!]
	\centering
	\includegraphics[width=\columnwidth]{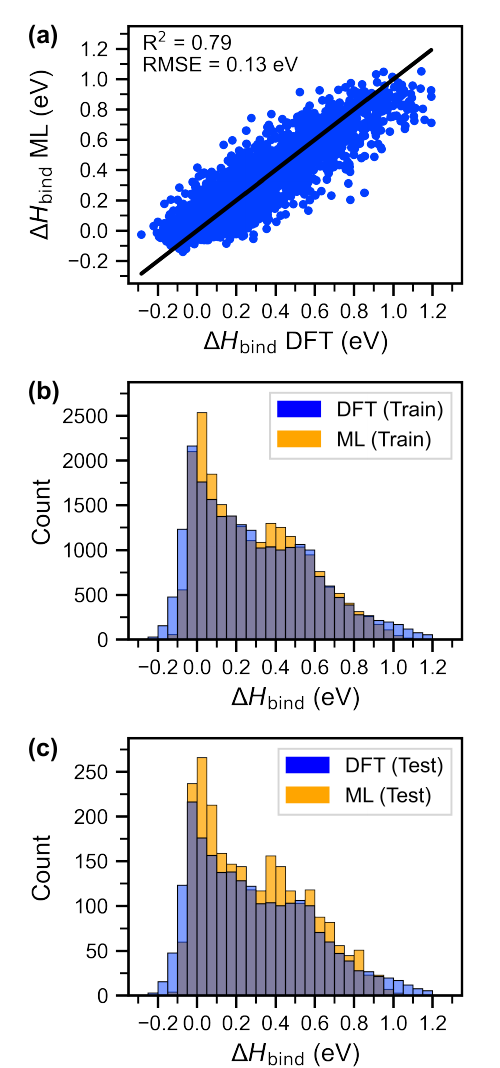}
	\caption[]{\label{fig:ml_results_2D} \small
		Performance of the best ML model.~(a) Truth plot for CO$_{2}$ binding enthalpies predicted by our best model versus DFT enthalpies for the amine binding sites in the testing set. The black line represents an ideal 1:1 linear correlation.~(b) Histogram of CO$_{2}$ binding enthalpies for DFT and ML model for the amine binding sites in the training set.~(c) Histogram of binding enthalpies for DFT and ML for the testing set.}
\end{figure}

Selected statistics for our DFT dataset used for training our ML models are given in the supplementary information (Figures~\ref{si-fig:statistics} and~\ref{si-fig:sa_and_ra_nist}).
Figure~\ref{fig:ml_results_2D} displays the performance of our most successful purely SMILES-based ML model.
The truth plot (Figure~\ref{fig:ml_results_2D}a) compares the enthalpies predicted by the ML model to the DFT level binding enthalpies for the molecules in the testing set.
The model has a coefficient of determination $R^{2}$ = 0.79 and a root mean squared error (RMSE) of 0.13 eV, with an RMSE of 0.08 eV for the region below 0 eV.
Histograms of the ML predicted and DFT computed binding enthalpies (Figures~\ref{fig:ml_results_2D}b,c) show overall similar distributions for both the training and testing sets.
The distribution is slightly compressed for the ML predictions compared to the DFT case, which is sensible given the low availability of data-points to train the model at the extremes of the enthalpy distribution.
Overall, our model's accuracy is sufficient to select promising active sites from the wide chemical space of the GDB-17 database.

\begin{figure*}[ht!]
	\centering
	\includegraphics[width=\textwidth]{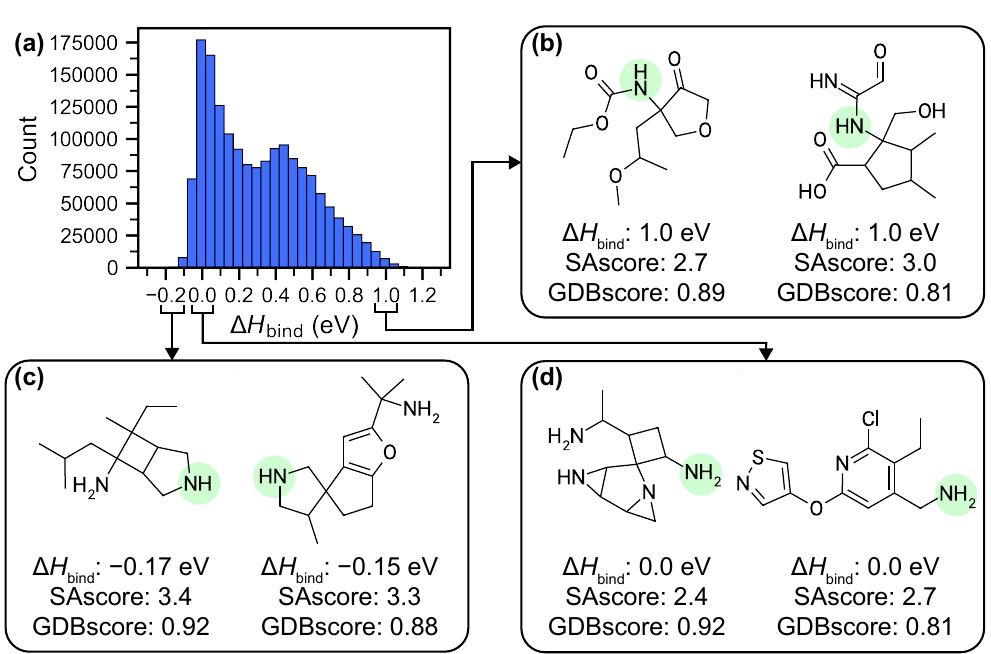}
	\caption{\label{fig:gdb_17_best_molecules}\small Results of high-throughput screening of $\sim$1.6 million molecules from the GDB-17 database.~(a) Distribution of binding energies in our GDB-17 set predicted by our best trained ML model. Representative molecules from (b) high $\Delta H_{\mathrm{bind}}$,~(c) low $\Delta H_{\mathrm{bind}}$, and (d) $\Delta H_{\mathrm{bind}}$ $\approx$ 0 eV regions.
		Molecules have been chosen from their respective regions after filtering based on synthesizability metrics (GDBscore > 0.64 and SAscore < 3.4).
		Functional groups corresponding to the given values are highlighted in green.
		The full set of molecules screened in the GDB-17 database together with predicted CO$_{2}$ binding enthalpies and synthesizability scores are available as supplementary information.}
\end{figure*}

To anchor computed DFT values in physical meaning, we also applied our methodology for computing binding enthalpies to a simplified branched polyethylenimine (BPEI) model (Figure~\ref{si-fig:bpei_si}), which is one of the most commonly used CO$_{2}$ sorbent materials.\cite{Zeeshan2023,Shi2020,Sanz-Perez2016}
In this case we obtain $\Delta H_{\mathrm{bind}}$ = -0.04 eV, giving us a benchmark for identifying structures which will bind CO$_{2}$ more or less tightly than BPEI\@.
Thus, we will be able to identify materials with enhanced CO$_2$ adsorption capacity and/or regenerative performance.
Ideal binding ranges are application dependent and based on complex engineering factors specific to given materials (e.g., target heat of regeneration)\cite{Zeeshan2023} which we refrain from discussing further.
Nevertheless, the models and methodology presented here will enable future endeavors in identifying molecular motifs within a given target CO$_2$ binding range that may be embedded in carbonaceous polymers.

We also trained a model utilizing 3D differential descriptors (requiring an optimized 3D geometry), information on which is available in the SI\@.
This model outperforms the models trained only on 2D, SMILES-derived descriptors and has an \emph{R$^{2}$} value of 0.87 and an RMSE of 0.10 eV (Figure~\ref{si-fig:ml_results_3d}) in the testing set.
However, this comes at the cost of needing a DFT-optimized geometry for both the ``parent'' and ``child'' molecule in order to compute the 3D descriptors required as input for the model.
Such an approach makes the ML approach redundant, as only thermodynamic corrections would then be needed to obtain binding enthalpies.
However, this 3D model provides a benchmark for the accuracy of our significantly faster and more practical 2D model by illustrating what can be achieved with a descriptor-based ML approach.

\subsection{Predictions by the Best ML Model}

\begin{figure}[ht!]
	\centering
	\includegraphics[]{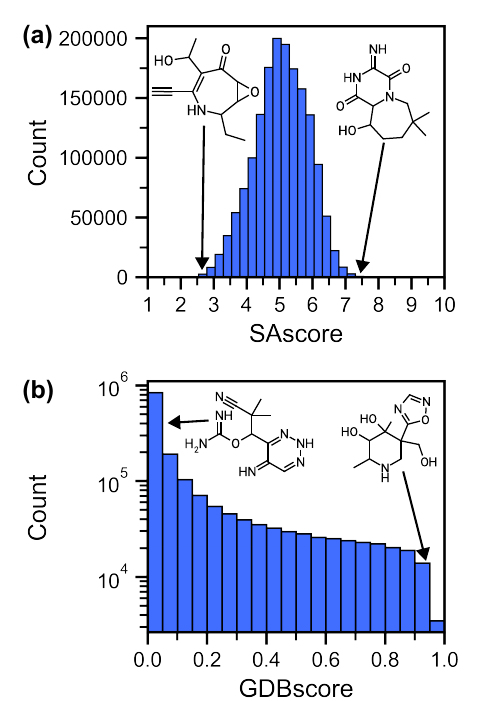}
	\caption[]{\label{fig:sa_and_ra_gdb17} \small Distributions of synthesizability scores for GDB-17 set.~(a) SAscore distribution.~(b) GDBscore distribution. Insets show example molecules selected from most unfavorable and favorable portions of the respective distributions.}
\end{figure}

Figure~\ref{fig:gdb_17_best_molecules}a displays a histogram of binding enthalpies predicted for approximately 1.6 million nitrogen binding sites from the GDB-17 database by our best 2D model.
Figures~\ref{fig:gdb_17_best_molecules}b, c, and d show representative molecules with favorable synthesizability scores chosen at random from high, low and intermediate regions of the distribution, respectively.
Immediately, it can be seen that the representative molecules from these three regions contain of distinct types of active sites (carbamate esters, secondary amines, and primary amines, respectively).
Figure~\ref{fig:sa_and_ra_gdb17} shows the distribution of SAscore and GDBscore synthesizability metrics for our dataset, showing that most molecules may require challenging synthetic approaches.
The SAscore shows a roughly normal distribution centered at 5, whereas the GDBscore predicts low probability of finding a synthetic route for most molecules.
This is sensible, as the GDB-17 has a large proportion of novel chemistries such as fused aromatic rings.\cite{Ruddigkeit2012}
There is a larger number of favorable synthesizability scores in the DFT set than in the larger GDB-17 set, due to the explicit inclusion of NIST molecules in the former (see Methods), slightly biasing the DFT set towards realizable materials.
However, this is not expected to significantly hamper predictive performance.
Figure~\ref{si-fig:best_mols_matrix} displays the 20 molecules with the most thermodynamically favorable CO$_{2}$ binding enthalpies which are also predicted to be synthesizable.
The full set of predictions and synthesizability scores is available as supplementary information.
The novel chemistries identified by this dataset are thus expected to accelerate design of efficient direct air capture materials.

\begin{figure}[ht!]
	\centering
	\includegraphics[]{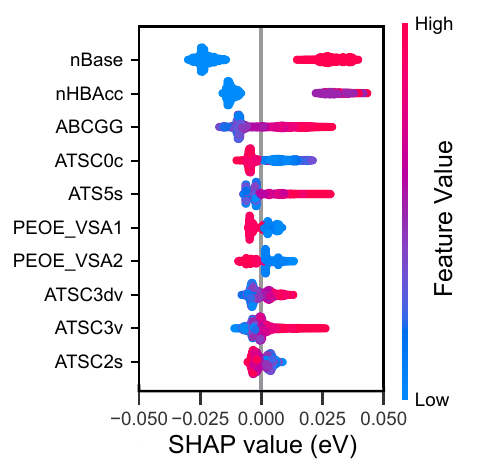}
	\caption{\label{fig:2DGBT_shap}\small Beeswarm plot of SHAP values\cite{Lundberg2020,SHAP} for the top 10 most important descriptors for determining CO$_{2}$ binding enthalpies in nitrogen-functionalized materials. SHAP values represent the influence of a feature's value (indicated by the heatmap) on the predicted binding enthalpy for a given input molecule. See Table~\ref{si-tab:beeswarm_descriptions} in the supplementary information for full definitions of each descriptor.}
\end{figure}

Nearly 11\% of molecules in our GDB-17 set have active sites that bind more strongly than BPEI and are therefore likely to have similar or higher CO$_{2}$ capacities.
Of these molecules, 2,642 have a GDBscore above 0.64 and an SAscore below 3.4.
These cutoff criteria are the mean plus (for GDBscore) or minus (for SAscore) two times the standard deviation for the GDB-17 set.
Of these binding sites, 1,887 are primary amines while 755 are secondary amines.
The absence of other N-containing functional groups suggests amines have the greatest affinity for CO$_{2}$.
Although secondary amines have been suggested to have more favorable adsorption characteristics,\cite{Gray2005,Zelenak2008} partially due to increased basicity resulting in greater affinity for CO$_{2}$, our results indicate that both primary and secondary amines can act as effective active sites, and that the extended structure of the active site can matter more than the specific functional group identity.
We further draw insight based on Tanimoto similarity scores (Figure~\ref{si-fig:tanimoto_matrix}), generally considered reliable for identifying closely related chemical structures,\cite{Rogers2010,Bajusz2015}.
For the 20 molecules with the strongest CO$_{2}$ binding enthalpies which are synthesizable, the mean Tanimoto similarity score is 0.15 (where Tanimoto similarity values can range from 0 to 1).
This indicates that the best CO$_{2}$ binders display significant heterogeneity.
Importantly, this analysis implies that discovery of amines with favorable CO$_{2}$ binding energetics is not trivial, and that a generic amine is unlikely to bind CO$_{2}$.

We further evaluate relative feature importance using a beeswarm plot of SHAP values (Figure~\ref{fig:2DGBT_shap}) for the training set.
Table~\ref{si-tab:beeswarm_descriptions} contains the descriptions from the Mordred documentation for the ten descriptors in Figure~\ref{fig:2DGBT_shap}.
The reader should keep in mind that all descriptors are differential descriptors, i.e., nBase in our model represents the \emph{change} in the number of bases upon CO$_{2}$ binding, rather than the absolute number of bases.
Some of these differential descriptors are chemically intuitive, such as nBase or nHBAcc (change in the number of hydrogen bond acceptors).
The ABCGG descriptor is an atom-bond connectivity descriptor, which describes the topology of the molecule.
Its importance according to SHAP values implies that the way the topology changes upon CO$_{2}$ binding is relevant, but further specific detail cannot easily be gleaned from the numerical value of this descriptor.
The same can be said of other topologically related descriptors, such as ATSC3v, which is an autocorrelation descriptor weighted by van der Waals volume and indicates how the van der Waals radii change upon CO$_{2}$ binding is important.
Although there is information loss in the construction of differential descriptors (i.e., the model does not know how many bases were in the original molecule) and they tell us less about the original molecule than conventional descriptors would, differential descriptors exhibited markedly superior performance to other approaches (Table~\ref{si-tab:desc_performance_table}).
Importantly, this implies that complex approaches such as our current ML strategy are necessary to capture the nuanced factors which lead towards favorable CO$_{2}$ binding performance.

\begin{figure*}[ht!]
	\centering
	\includegraphics[]{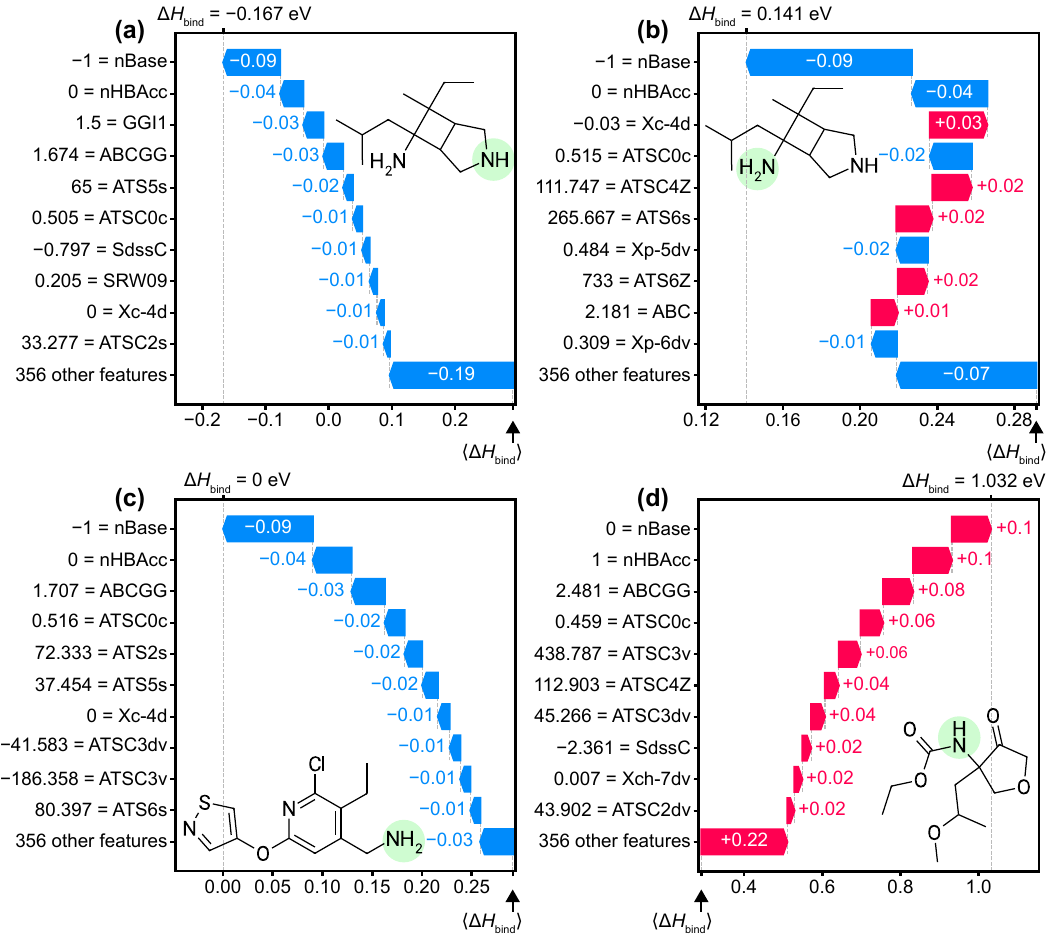}
	\caption{\label{fig:shap_waterfall_and_mols}\small SHAP waterfall plots for our best ML model for selected structures from Figure \ref{fig:gdb_17_best_molecules}. Each plot shows the contribution of features in moving from the mean CO$_{2}$ binding enthalpy of the dataset, $\langle \Delta H_{\mathrm{bind}} \rangle$, to the final predicted binding enthalpy of the considered molecule. The specific values for each descriptor for the corresponding inset molecule are given on the y axis. Waterfall plots for (a) a binding site with a $\Delta H_{\mathrm{bind}}$ significantly below our BPEI benchmark of -0.04 eV,~(b) the other binding site on the same molecule in (a),~(c) a binding site with a final $\Delta H_{\mathrm{bind}}$ value of 0, and (d) a binding site with a $\Delta H_{\mathrm{bind}}$ from the upper end of the distribution. The amine binding site for each plot is highlighted in green on the inset line angle structure of the given molecule.}
\end{figure*}

Figure~\ref{fig:shap_waterfall_and_mols} shows SHAP waterfall plots, which allow us to see the influence of each descriptor in the model's decision-making for individual molecules selected from Figure 3 to represent distinct regions of binding enthalpy.
The nBase differential descriptor has a value of ``-1'' for primary and secondary amines for subfigures~\ref{fig:shap_waterfall_and_mols}a, b, and c, but has a value of ``0'' for the carbamate ester in subfigure~\ref{fig:shap_waterfall_and_mols}d.
Similarly, the nHBAcc descriptor denotes a change in the number of hydrogen bond acceptors.
Thus, these descriptors discriminate between different functional group active sites.
Although other nitrogen-based moieties have been investigated for direct air capture\cite{Petrovic2021}, primary and secondary amines are the most common, and our results suggest that they tend to have stronger binding enthalpies due to their increased basicity over other functional groups.
Figures~\ref{fig:shap_waterfall_and_mols}a and b highlight different nitrogen active sites on the same molecule that lead to drastically different binding enthalpies (-0.167 eV vs 0.141 eV), showing that our model captures local effects specific to each binding site, in additional to global contributions from the overall structure of the molecule.
Lastly, the mean absolute value of Spearman correlations for descriptors in our input vector (Figure~\ref{si-fig:descriptors_corr}) is 0.27 in the training set.
This suggests that the majority of these descriptors capture unique facets of a given molecule.

Overall, the SHAP values for each descriptor are small (<0.1 eV), and none of them dominate the model's decision-making with regard to binding enthalpy, suggesting difficulty in using a small number of descriptors as a simple guideline for rational design of CO$_{2}$ adsorption sites.
Thus, while the descriptor approach undertaken here allows for model interpretability and some chemical insight, the conclusion suggested by the present work is that \emph{no one particular molecular feature makes an overwhelming contribution to the CO$_2$ binding enthalpy}.
Such complex interplay of molecular features necessitates the complexities of an ML model trained on DFT data in order to effectively explore such a large chemical space, and active sites cannot be chosen for direct air capture with simple heuristics.

%%% Local Variables:
%%% mode: latex
%%% TeX-master: "../main"
%%% End:

\section{Discussion}
The binding enthalpies for over 1.6 million potentially active sites for direct air capture materials reported here should be critical in advancing the state of the art for direct air capture.
Specifically, 2,646 amines are predicted to both be synthesizable and to have more favorable binding enthalpies than BPEI\@.
The novel structures uncovered in this work will design direct air capture materials with optimized capacity and/or heat of regeneration.
As our model interpretability techniques suggest that amines bind CO$_{2}$ more tightly than other N-bearing functional groups, future materials discovery campaigns for direct air capture would be best focused on amines over other N-bearing groups.
However, our results indicate that both primary and secondary amines can be effective binding sites, and other optimization considerations (e.g., stability) will influence their relative utility.
Examining stability, for example to oxygen, for the dataset reported herein is a reasonable future improvement for the current dataset.
The binding enthalpies predicted herein could conceivably be used to parameterize an adsorption isotherm model for a material incorporating these active sites\cite{Young2021,Liu2024}.
Such an approach would offer a tantalizing way to predict materials performance from atomistic DFT calculations and could play a key role in uncovering transformational direct air capture materials.

Additionally, our interpretability techniques also suggest that CO$_{2}$ chemisorption is a complex phenomenon and that discovery of materials for direct air capture is not trivial.
In particular, developing heuristics for selecting strong CO$_{2}$ binding sites seems impractical, as amines with similar predicted binding enthalpy have very low similarity, and there are no input features which make overwhelming contributions to predicted binding energetics.
As the qualities which contribute to CO$_{2}$ adsorption are shown to be subtle and nuanced, ML approaches as employed in this work are critical for exploring the diverse chemical space available for these materials and selecting chemistries with improved properties for direct air capture.
We have provided the tools to train these models, as well as the full set of data predicted by our model for use in attempts to optimize CO$_{2}$ direct air capture polymers.
The full set of DFT enthalpies, ML predictions and the means to reproduce our ML models are also made available as supplementary material.
An extension of the current approach to use descriptors depending on 3D structural information, possibly from a computationally inexpensive force-field or semi-empirical optimization, may be a viable improvement over the purely SMILES-based approach presented here.

Looking forward, we expect the active site candidates identified herein to be used as guidance for development of more efficient direct air capture materials.
Furthermore, the novel methodology outlined in this work, such as the utilization of the differential descriptors defined herein, will be useful for other efforts where binding energetics are key modeling targets, such as in the understanding of catalytic pathways and intermediates.
These advances should be critical in advancing the state of the art of direct air capture, increasing the efficiency of these devices and thereby helping to build the road towards a more economically robust green energy future.

%%% Local Variables:
%%% mode: latex
%%% TeX-master: "../main"
%%% End:

\section{Methods}
The initial step of our workflow (Figure~\ref{fig:workflow}) is the selection of input molecules.
16,334 molecules containing at least one primary or secondary-substituted nitrogen-bearing moiety are selected at random from the publicly available GDB-17 database together with 627 molecules with entries in the NIST database which fit the GDB-17 criteria.\cite{Ruddigkeit2012}
The selection criteria includes primary and secondary amines as well as other N-containing functional groups (e.g., imines) so that more diverse chemistries may be explored, as other nitrogen-based moieties can exhibit CO$_{2}$-philic properties.\cite{Petrovic2021}
Table~\ref{si-tab:functional_groups} details the functional group composition of this dataset.
These are dubbed ``parent'' molecules.
998 of these molecules either failed to converge with DFT computations, or the derived ``child'' molecules (described below) failed to converge.
This leaves 15,336 total parent molecules for our DFT dataset, 620 of which have entries in the NIST database.
Some of these molecules have more than one active site moiety, resulting in a total of 24,819 active sites.

These molecules are collected as SMILES (Simplified Molecular Input Line Entry System) representations,\cite{Weininger1988} which describe the 2D connectivity graph of the given molecule in textual form.
Three-dimensional conformations of these molecules are generated using the ETKDG method\cite{Riniker2015} implemented in RDKit.\cite{RDKit}
The number of conformations generated is based on the number of rotatable bonds following the heuristic given by Ebejer et al.\cite{Ebejer_2012}
Subsequently, GFN2-xTB,\cite{Bannwarth2019} a semi-empirical density functional tight binding method which can produce accurate geometries,\cite{Menzel2021} is used to relax these initial conformations and select the minimum energy conformer.
DFT geometry optimization is then performed using Gaussian 16 with the B3LYP functional and 6-311+G(d,p) basis set.\cite{g16,LYP86,Becke93,Clark1983,Spitznagel_1987,Krishnan1980,McLean1980}
Harmonic frequencies are then calculated at the minimized geometry, providing thermodynamic corrections to the electronic energy within the harmonic limit and confirming that the geometry is a true minimum on the potential energy surface.
Finally, we use these corrections to obtain enthalpies under standard temperature and pressure for each molecule.
All DFT computations, collection, and storage of properties are managed using the pyiron framework.\cite{Janssen2019}

For this work we investigate dry adsorption conditions, as they are more amenable to our high-throughput approach, owing to the added complexity of reaction mechanisms involving interaction with water molecules.\cite{Said2020}
Additionally, we are motivated to discover materials for dry CO$_{2}$ capture as humidity plays an important role in degradation of amine-based sorbents\cite{Carneiro2023} and can significantly increase energy requirements for regeneration of the material.\cite{Quang2014}
Modeling of binding energetics of polymers is a complex endeavor, and numerous contributions to the binding energy may be modeled with DFT including long-range polymeric effects in addition to energetics of bond formation.\cite{Glenna_2023}
Herein, we focus on the chemical contribution of the active site to the binding energy.
Dry conditions for CO$_{2}$ sorption primarily lead to ammonium carbamate formation\cite{Nguyen2023,Danon2011,Foo2016}.
Previous work shows that protonation of the bound CO$_{2}$ group is energetically favorable for the ammonium carbamate complex\cite{Lee2012,Li_2016} and that enthalpies of formation for carbamic acid and ammonium carbamate are closely correlated\cite{Lee2012}.
Figure~\ref{si-fig:dma_trend} shows that the stabilizing effect of an amine contributes a mean stabilization energy of -0.46 eV across the entire range of our dataset with a nearly linear correlation ($R^2$ = 0.99) to the enthalpy of the carbamic acid formation reaction.
Thus, we choose a carbamic acid formation mechanism for our high-throughput approach.
Once the DFT-relaxed structures are obtained, reaction SMARTS (SMILES Arbitrary Target Specification)\cite{SMARTS}  are used to transform active sites into their corresponding carbamic acid groups (``child'' molecules), following the scheme shown below for primary and secondary nitrogen sites:
\begin{subequations}
	\begin{equation}
		\ch{R-NH2 + CO2 -> R-NHCO2H,}
	\end{equation}
	\begin{equation}
		\ch{R2-NH + CO2 -> R2-NCO2H.}
	\end{equation}
\end{subequations}

3D conformations of these ``child'' molecules are obtained by allowing the coordinates of the bound CO$_{2}$H group to vary, while fixing the coordinates of all other atoms to those of the DFT-optimized structure of the parent molecule.
These conformations are then relaxed by GFN2-xTB in the same manner as for the parent molecules.
This is done to avoid meta-stable configurations by efficiently screening through possible torsional angles of the CO$_{2}$H group.
These ``child'' structures are then optimized using the B3LYP/6-311+G(d,p) level of theory as implemented in Gaussian 16, followed by a frequency calculation.

We next compute the change in enthalpy upon CO$_{2}$ binding according to the following scheme for primary and secondary sites, respectively:
\begin{subequations}
	\begin{equation}
		\Delta H_{\mathrm{bind}} = H_{\mathrm{R-NHCO_{2}H}} - H_{\mathrm{R-NH2}} - H_{\mathrm{CO_{2}}},
	\end{equation}
	\begin{equation}
		\Delta H_{\mathrm{bind}} = H_{\mathrm{R_{2}-NCO_{2}H}} - H_{\mathrm{R_{2}-NH}} - H_{\mathrm{CO_{2}}}.
	\end{equation}
\end{subequations}
Utilizing our automated high-throughput workflow, we collect binding enthalpies for 24,819 ``children'' corresponding to all possible nitrogen active sites in the ``parent'' molecules.
These binding enthalpies are used to train the machine learning surrogate models, which learn to predict CO$_{2}$ binding enthalpies from descriptors of a given SMILES string.
It should be noted that one molecule may have multiple binding sites, and we consider each possible reaction as a separate set of input data for training our model.
That is, the model is given all possible binding enthalpies for a given ``parent'' together with the corresponding ``child'', not just sites with the strongest binding enthalpy or the average over all binding sites in a molecule.

Next, the machine learning models are trained on the collected DFT dataset.
Our approach follows previous reports~\cite{Pinheiro2020} of neural networks trained using Mordred descriptors\cite{Moriwaki2018} derived from SMILES strings that achieved accurate prediction of seven molecular properties, including enthalpies of formation within 0.07 eV of DFT values.
Such an approach allows for tremendous increase in speed and versatility for high-throughput screening, as no 3D conformations of the molecules need to be generated.
The set of Mordred descriptors (\textbf{x}) computed for the molecules in our dataset is used as the input vectors for our models.\cite{Moriwaki2018}
Some descriptors were not able to be computed for all molecules in the test set.
Of the 1,826 descriptors available in Mordred (1,612 including only 2D descriptors), 1,484 remain after discarding those with missing values.
As an example, the ``MaxsLi'' descriptor, which relates to the electronic environment of the lithium atom, is missing for molecules that do not contain lithium.
Furthermore, because binding enthalpy depends on the energetics of the molecule before and after CO$_{2}$ adsorption occurs, there are multiple options for which set of Mordred descriptors should be used.
The approach used herein is to construct ``differential descriptors'', which is simply the difference in value for each descriptor between the ``child'' (CO$_{2}$ chemisorbed) and ``parent'' molecule:
\begin{equation}
	\Delta\textbf{x} = \textbf{x}_{\textrm{child}} - \textbf{x}_{\textrm{parent}}.
\end{equation}
Differential descriptors were found to provide superior predictions of the binding enthalpy compared to merely training the model with both sets of ``parent'' and ``child'' descriptors, and no improvement was seen from training on all three sets of descriptors compared to using differential descriptors only (see Table~\ref{si-tab:desc_performance_table}).
The descriptors are further down-selected by removing those with a Spearman correlation coefficient greater than 0.85 with any other descriptor (see Table~\ref{si-tab:correlated_2D_GBT}).
Several ML regression models implemented in scikit-learn\cite{scikit-learn} were tested (see Table~\ref{si-tab:model_performance_table}) with gradient boosting trees (GBT)\cite{Friedman2002} exhibiting the strongest performance.

The models are trained using a train-test split of 90:10 constructed from the DFT dataset outlined above.
The training set is further split into cross-validation sets for optimization of hyperparameters (see Table~\ref{si-tab:hyperparameter_table}) with a 90:10 training:validation ratio.
Once the hyperparameters are chosen, the model is fit on the full set of training data and the test set is used to validate model accuracy.
Uncertainty quantification is performed by using an ensemble of 150 regressors, which are trained on bootstrapped subsets of the training data equal to one half of the original dataset and generated through random sampling with replacement.
This ensemble model is the final one used for collecting performance metrics on the testing set and screening applications.
SHAP scores are used to assess feature importance in an attempt to provide meaningful chemical insight from these models in addition to their usage in high-throughput screening.\cite{Lundberg2020,SHAP}

Finally, the trained models are used to predict the binding enthalpies for nitrogen-bearing molecules in the GDB-17 database that were not present in the original dataset.
Binding enthalpies for 992,959 ``parent'' molecules are predicted using our best model for each binding site, totaling to 1,650,601 predicted binding enthalpies.
We also use the SAscore\cite{Ertl_2009} and GDBscore\cite{Thakkar_2021} metrics to estimate synthesizability of molecules for both our DFT dataset and the GDB-17 set.
The former metric is a simple heuristic model based on features such as the number of fused rings and other complex substructures, giving a value between 1 and 10, where a lower number represents a molecule that is more likely to be synthesizable.
The latter one is a machine learning surrogate model, trained on GDB molecules, which predicts whether the retrosynthetic analysis program AiZynthFinder\cite{Genheden2020} can identify a viable synthetic route for a given molecule.
The GDBscore is intended for classification and ranges from 0 to 1, with a higher number representing a higher probability that a given molecule is synthesizable.

%%% Local Variables:
%%% mode: latex
%%% TeX-master: "../main"
%%% End:


\nolinenumbers

\begin{acknowledgement}
	This research used resources provided by the Los Alamos National Laboratory Institutional Computing Program, which is supported by the U.S. Department of Energy National Nuclear Security Administration under Contract No. 89233218CNA000001.
	The authors acknowledge the Laboratory Directed Research and Development program of Los Alamos National Laboratory under Project No. 20230065DR.
	MCD also thanks the Center for Nonlinear Studies at LANL for financial support under Project No. 20220546CR-NLS.
	The authors acknowledge Harshul Thakkar and Rajinder Singh for discussions on direct air capture, and Navneet Goswami for discussions on isotherm models.
	We also acknowledge Jan Janssen and Michael G. Taylor for assistance with pyiron and workflow development.
	Benjamin Stein and Cassandra Gates are acknowledged for discussion on modeling branched polyethylenimine.
\end{acknowledgement}

\begin{suppinfo}

	The following files are available free of charge.
	\begin{itemize}
		\item Supplementary Information: Selected statistics for DFT dataset, detailed analysis of machine learning models and discovered amines (PDF).
		\item structures.zip: B3LYP/6-311+G(d,p) energies, structures in .xyz format, and synthesizability scores for DFT dataset and BPEI model. ML predicted energies and synthesizability scores for GDB-17-amine set.
	\end{itemize}

\end{suppinfo}

% \bibliography{/Users/megand/bib/references.bib}
\bibliography{references.bib}

\end{document}

\clearpage
\section{Mordred descriptor definitions}
\begin{table*}[ht!]
	\caption[]{\label{tab:beeswarm_descriptions}\small Definitions from Mordred documentation\cite{Moriwaki2018} for descriptors shown in Fig. 5}
	% \vspace{4mm}
	\small
	\begin{tabular}[tb]{cl}
		\toprule
		Descriptor & Definition                                                                   \\
		\midrule
		nBase      & basic group count                                                            \\
		nHBAcc     & number of Hydrogen bond acceptors                                            \\
		ABCGG      & Graovac-Ghorbani atom-bond connectivity index                                \\
		ATSC0c     & centered Moreau-Broto autocorrelation of lag 0 weighted by Gasteiger charge  \\
		ATS5s      & Moreau-Broto autocorrelation of lag 5 weighted by intrinsic state            \\
		PEOE\_VSA1 & Molecular Operating Environment (MOE) Charge                                 \\ & van der Waals surface area Descriptor 1 \\
		PEOE\_VSA2 & Molecular Operating Environment (MOE) Charge                                 \\ & van der Waals surface area Descriptor 2 \\
		ATSC3dv    & centered Moreau-Broto autocorrelation of lag 3 weighted by valence electrons \\
		ATSC3v     & centered Moreau-Broto autocorrelation of lag 3 weighted                      \\ & by van der Waals volume \\
		ATSC2s     & centered Moreau-Broto autocorrelation of lag 2 weighted by intrinsic state   \\
		\bottomrule
	\end{tabular}

\end{table*}

Table~\ref{tab:beeswarm_descriptions} gives definitions from the Mordred documentation for the descriptors in Figure 5 of the main text.

\clearpage
\subsection{Functional Groups in Dataset}

Table~\ref{tab:functional_groups} gives information on the SMARTS strings used for creating our DFT dataset together with SMARTS strings for identifying functional groups.
SMARTS are a chemical language, similar to SMILES, for matching molecular patterns.\cite{SMARTS}
We identify the functional groups for active sites using RDKit by finding all molecules which match the given SMARTS, and then checking if the match includes the specific atoms of a given active site.
By doing so, we can identify the functional group identity of each active site in an impartial, high-throughput manner.
We used the given ``Primary amine (Expanded)'' and ``Secondary amine (Expanded)'' definitions to construct our dataset.
As mentioned in the main text, we used this criteria in order to explore more novel and diverse chemistries for direct air capture materials.
The other given SMARTS matches allow us to identify the specific functional group identities present as active sites in our dataset.
SMARTS for nitrogen-containing functional groups that resulted in zero matches for our active sites are omitted.

\begin{table}[htbp]
	\caption[]{\label{tab:functional_groups} Number of active sites by functional group in the DFT dataset evaluated using the given SMARTS string. For primary and secondary amines, two definitions are included: an expanded definition that was used to construct the DFT dataset, which matches amides and other closely related functional groups, and a narrow definition from the Daylight webpage\cite{SMARTS} which is more restrictive to formal amine groups.}
	\vspace{4mm}
	\small
	\begin{tabular}[tb]{llr}
		\toprule
		Functional group           & SMARTS                                                  & \# in Dataset \\
		\midrule
		%  primary\_or\_secondary & [NX3;H2,H1;!\$(NC=O)]                                                             & 21186 \\
		Primary amine (Expanded)   & [!O;*:2]-[Nv3\&H2:1]                                    & 11387         \\
		Secondary amine (Expanded) & ([Nv3\&H1:1](-[!O;\*:2])-[!O;\*:3]),[nv3\&H1:1]         & 13423         \\
		Primary amine (Daylight)   & [NX3;H2;!\$(NC=[!\#6]);!\$(NC\#[!\#6])][\#6]            & 8935          \\
		Secondary amine (Daylight) & [NX3;H1;!\$(NC=[!\#6]);!\$(NC\#[!\#6])][\#6]            & 7601          \\
		%   two\_amines          &   [NX3;H2,H1;!\$(NC=O)].[NX3;H2,H1;!\$(NC=O)]                                     & 13009 \\
		Enamine                    & [NX3][CX3]=[CX3]                                        & 396           \\
		Enamine or aniline         & [NX3][\$(C=C),\$(cc)]                                   & 2535          \\
		Amino acid                 & [NX3,NX4+][CX4H]([*])[CX3](=[OX1])[O,N]                 & 323           \\
		Azole                      & [\$([nr5]:[nr5,or5,sr5]),\$([nr5]:[cr5]:[nr5,or5,sr5])] & 784           \\
		Hydrazine                  & [NX3][NX3]                                              & 27            \\
		Hydrazone                  & [NX3][NX2]=[*]                                          & 8             \\
		Amide                      & [NX3][CX3](=[OX1])[\#6]                                 & 1428          \\
		% Se &   [nv3\&H1:1]                                                                     & 1495 \\
		\bottomrule
	\end{tabular}

\end{table}

\clearpage

\subsection{ML Model Training}

Tables~\ref{tab:desc_performance_table}-\ref{tab:hyperparameter_table} contain information on the various machine learning models examined.
All models were trained using Scikit-learn 1.3.0.
As mentioned in the main text, uncertainty quantification for all models was performed using bootstrapped ensemble models, consisting of 150 estimators trained on datasets one half the size of the original dataset generated through random sampling with replacement.
For RMSE and \emph{R$^{2}$}, uncertainty is given as the standard deviation from the ensemble model multiplied by a factor of 1.96, representing a 95\% confidence interval.
The machine learning models investigated were Gradient Boosted Trees (GBT), K-Nearest Neighbors (KNN), Random Forest (RF), Artificial Neural Networks (ANN) and Linear regression.
Hyperparameters were optimized for each model discussed, using the parameters in Table~\ref{tab:hyperparameter_table} for a halving grid search.
The grid searches were performed using a 90:10 train:cross-validation split on the training set with 20 cross-validation sets.
The GBT model performs the best, although RF is close in performance with an \emph{R$^{2}$} of 0.78.
Surprisingly, the Linear model performs as well as the RF model and outperforms both KNN and ANN\@.
Table~\ref{tab:desc_performance_table} gives the comparison of GBT models trained on different descriptor sets.
For each set of descriptors, the Spearman correlation matrix used to remove redundant descriptors was computed uniquely.
For example, for ``Parent + Child + Differential'', if a pair of descriptors has a correlation coefficient greater than 0.85, one of the descriptors was removed from the input vector used for training the model (See Table~\ref{tab:correlated_2D_GBT}).
Table~\ref{tab:model_performance_table} contains the descriptions from the Mordred documentation for the ten descriptors in Figure X of the main text.

\begin{table}[htbp]
\caption[]{\label{tab:desc_performance_table} R$^{2}$ and RMSE (in eV) in the testing set for an optimized gradient boosting trees model trained on each given set of descriptors. Hyperparameters for each model are optimized using the grid from Table \ref{tab:hyperparameter_table}.}
\vspace{4mm}
\begin{tabular}[tb]{lrr}
\toprule
%\multicolumn{2}{c}{GBT} \\
Descriptor set & RMSE (eV) & R$^{2}$ \\
\midrule
Differential & 0.132 $\pm$ 0.002 & 0.794 $\pm$ 0.006 \\
Parent & 0.227 $\pm$ 0.001 & 0.398 $\pm$ 0.007 \\
Child & 0.156 $\pm$ 0.002 & 0.713 $\pm$ 0.008 \\
Parent + Child & 0.151 $\pm$ 0.002 & 0.731 $\pm$ 0.006 \\
Parent + Child + Differential & 0.132 $\pm$ 0.002 & 0.796 $\pm$ 0.005 \\
\bottomrule
\end{tabular}

\end{table}

\begin{table}[htbp]
	\caption[]{\label{tab:correlated_2D_GBT} Differential descriptors which were highly correlated (Spearman correlation coefficient >0.85) with the top 20 differential descriptors by SHAP value from the best 2D ML model. Highly correlated descriptors were not used for training ML models. The full list of correlated descriptors for the best 2D model are available as an additional file.}
	\vspace{4mm}
	\begin{tabular}[tb]{cr}
		\toprule
		Base descriptor & Correlated descriptors                \\
		\midrule
		nBase           & ATSC1c                                \\
		nHBAcc          & ---                                   \\
		ABCGG           & SpMAD\_Dzse, SpMAD\_D, AETA\_eta\_R   \\
		ATSC0c          & ---                                   \\
		ATS5s           & ATS4v, ATS5se, ATS5pe, ATS5are, ATS4p \\
		PEOE\_VSA1      & ---                                   \\
		PEOE\_VSA2      & ---                                   \\
		ATSC3dv         & AATSC3dv, MATS3dv                     \\
		ATSC3v          & AATSC3v, MATS3v, GATS3v               \\
		ATSC2s          & ---                                   \\
		ATS6s           & ATS6se, ATS6pe, ATS6are               \\
		Xch-7dv         & ---                                   \\
		Xc-4d           & Xc-4dv                                \\
		SRW09           & ---                                   \\
		SMR\_VSA6       & ---                                   \\
		ATS2s           & ATS4s, ATS3Z, ATS3m                   \\
		SdssC           & ---                                   \\
		Xp-7dv          & ---                                   \\
		MID\_h          & ---                                   \\
		AATS4d          & AATS4v, AATS4p                        \\
		\bottomrule
	\end{tabular}

\end{table}

\begin{table}[htbp]
\caption[]{\label{tab:model_performance_table} R$^{2}$ and RMSE in eV in the testing set for each given optimized model. All models are trained on 2D differential descriptors. Hyperparameters for each model are optimized using the grid from Table \ref{tab:hyperparameter_table}.}
\vspace{4mm}
\begin{tabular}[tb]{lrr}
\toprule
%\multicolumn{2}{c}{GBT} \\
Model & RMSE (eV) & R$^{2}$ \\
\midrule
GBT & 0.132 $\pm$ 0.002 & 0.794 $\pm$ 0.006 \\
KNN & 0.148 $\pm$ 0.002 & 0.742 $\pm$ 0.007 \\
RF & 0.137 $\pm$ 0.001 & 0.780 $\pm$ 0.004 \\
ANN & 0.141 $\pm$ 0.004 & 0.767 $\pm$ 0.014 \\
Linear & 0.137 $\pm$ 0.002 & 0.780 $\pm$ 0.005 \\
\bottomrule
\end{tabular}

\end{table}

\begin{table}[htbp]
  \caption[]{\label{tab:hyperparameter_table} Table of hyperparameters optimized for each model. The best hyperparameters discovered for each model are bolded. All hyperparameters not explicitly included are left as the scikit-learn defaults. Hyperparameters were optimized using a Halving Grid Search with 20 cross-validation splits with a 90:10 train:validation ratio. The ``neg\_root\_mean\_squared\_error'' metric was chosen for scoring the grid search. For ANN, an initial full grid search was performed examining hidden layer sizes 128, 256, 512, 768 and 1024 and a number of hidden layers between 1 and 10. The value for hidden\_layer\_sizes used for the Halving Grid Search was the best found from this first search.}
\vspace{4mm}
\begin{tabular}[tb]{ccr}
  \toprule
  Model & Hyperparameter & Options \\
\midrule
GBT & learning\_rate & [0.01, \textbf{0.1}, 1] \\
& n\_estimators & [300, 600, \textbf{1000}] \\
& max\_depth & [\textbf{3}, 9, 12, None] \\
& min\_samples\_leaf & [1, \textbf{0.01}] \\
& subsample & [0.1, \textbf{1.0}] \\
& min\_impurity\_decrease & [0.0, \textbf{0.1}] \\
& max\_features & [\textbf{1.0}, ``sqrt'', ``log2''] \\
& random\_state & [\textbf{42}] \\
\midrule
KNN & n\_neighbors & [\textbf{20}, 300, 600] \\
& weights & [``uniform'', \textbf{``distance''}] \\
& algorithm & [\textbf{``auto''}, ``ball\_tree'', ``kd\_tree'', ``brute''] \\
& leaf\_size & [30, 70, \textbf{110}] \\
& p & [\textbf{1}, 2, 3] \\
\midrule
RF & n\_estimators & [25, 50, \textbf{100}] \\
& max\_depth & [\textbf{None}] \\
& criterion & [\textbf{``squared\_error''}] \\
& min\_samples\_split & [2, \textbf{4}, 8] \\
& min\_samples\_leaf & [\textbf{1}] \\
& max\_features & [\textbf{``sqrt''}] \\
& bootstrap & [\textbf{False}] \\
& min\_impurity\_decrease & [\textbf{0.0}] \\
& random\_state & [\textbf{42}] \\
\midrule
ANN & hidden\_layer\_sizes & [\textbf{(768, 768, 768, 768, 768, 768)}] \\
& activation & [\textbf{``relu''}] \\
& solver & [\textbf{``adam''}] \\
& alpha & [\textbf{0.001}, 0.01, 0.1] \\
& learning\_rate & [\textbf{``adapative''}] \\
& learning\_rate\_init & [\textbf{0.001}, 0.01] \\
& random\_state & [\textbf{42}] \\
& max\_iter & [\textbf{15000}] \\
\bottomrule
\end{tabular}

\end{table}

\clearpage

\subsection{Dimethylamine Stabilization}

\begin{figure*}[ht]
	\centering
	\includegraphics[]{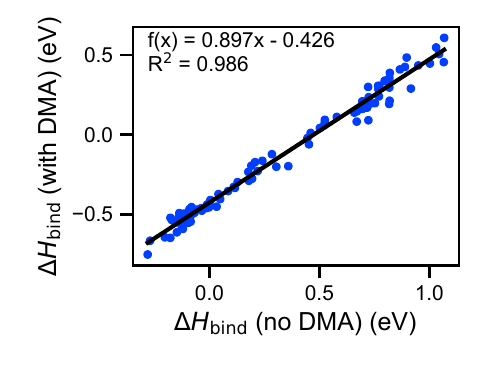}
	\caption[]{\label{fig:dma_trend} Binding enthalpies with an explicit dimethylamine added to form an ammonium carbamate complex according to equations \ref{eq:dma_rxn1} and \ref{eq:dma_rxn2} versus binding enthalpies computed using the simpler carbamate mechanism described in the main text.}
\end{figure*}

Figure~\ref{fig:dma_trend} shows the correlation between $\Delta H_{{\mathrm{bind}}}$ for 105 samples from our DFT dataset and for $\Delta H_{{\mathrm{bind}}}$ when an explicit dimethylamine (DMA) molecule is added to form an ammonium carbamate complex. These binding enthalpies are computed according to the following scheme:
\begin{subequations}
	\begin{equation}
		\ch{R-NH2 + CO2 + DMA -> R-NHCO2H--NH(CH3)2,}
		\label{eq:dma_rxn1}
	\end{equation}
	\begin{equation}
		\ch{R2-NH + CO2 + DMA-> R2-NCO2H--NH(CH3)2.}
		\label{eq:dma_rxn2}
	\end{equation}
\end{subequations}
The 3D structure for the ammonium carbamate complex is obtained by taking the optimized structure of the corresponding carbamic acid molecule from the original dataset and adding a DMA group coordinated via an intermolecular hydrogen bond from the carbamic acid group to the DMA\@.
Conformations are generated for the ammonium carbamate complex by allowing the DMA atom positions to change while freezing the coordinates of the carbamate atoms.
Conformers are relaxed with GFN2-xTB and then a B3LYP/6311+G(d,p) calculation is done on the lowest energy conformer to get the enthalpy of the complex.
$\Delta H_{\mathrm{bind}}$ for equations~\ref{eq:dma_rxn1}~and~\ref{eq:dma_rxn2} are then calculated.
For Figure~\ref{fig:dma_trend}, structures were selected at random from three distinct regions: low binding enthalpy ($\Delta H_{\mathrm{bind}}$ < -0.074 eV), intermediate binding enthalpy (-0.074 eV < $\Delta H_{\mathrm{bind}}$ < 0.658 eV), and high binding enthalpy ($\Delta H_{\mathrm{bind}}$ > 0.658 eV).
These cutoffs are the mean $\Delta H_{\mathrm{bind}}$ in the DFT dataset $\pm$ 1.25 times the standard deviation.
The stabilization energy from DMA can then be easily determined:
\begin{equation}
	\Delta H_{\mathrm{stabilization}} = \Delta H_{\mathrm{bind}} \mathrm{(with~DMA)} - \Delta H_{\mathrm{bind}} \mathrm{(no~DMA)}.
	\label{eq:stable}
\end{equation}
The stabilization energies for each region are -0.411 $\pm$ 0.067 eV, -0.444 $\pm$ 0.070 eV, and -0.514 $\pm$ 0.117 eV for the low, intermediate and high regions, respectively, with an overall mean stabilization of -0.456 eV.
The slope of the trendline is 0.897, resulting in the stabilization energy increasing slightly as $\Delta H_{\mathrm{bind}}$ increases.
A plausible explanation is that, since the carbamic acid group is less strongly bonded to the parent structure at higher binding enthalpies, the additional stabilization from the DMA group plays a more important in stabilizing the carbamic acid group.

\clearpage

\subsection{Dataset Statistics}

\begin{figure*}[ht]
	\centering
	\includegraphics[]{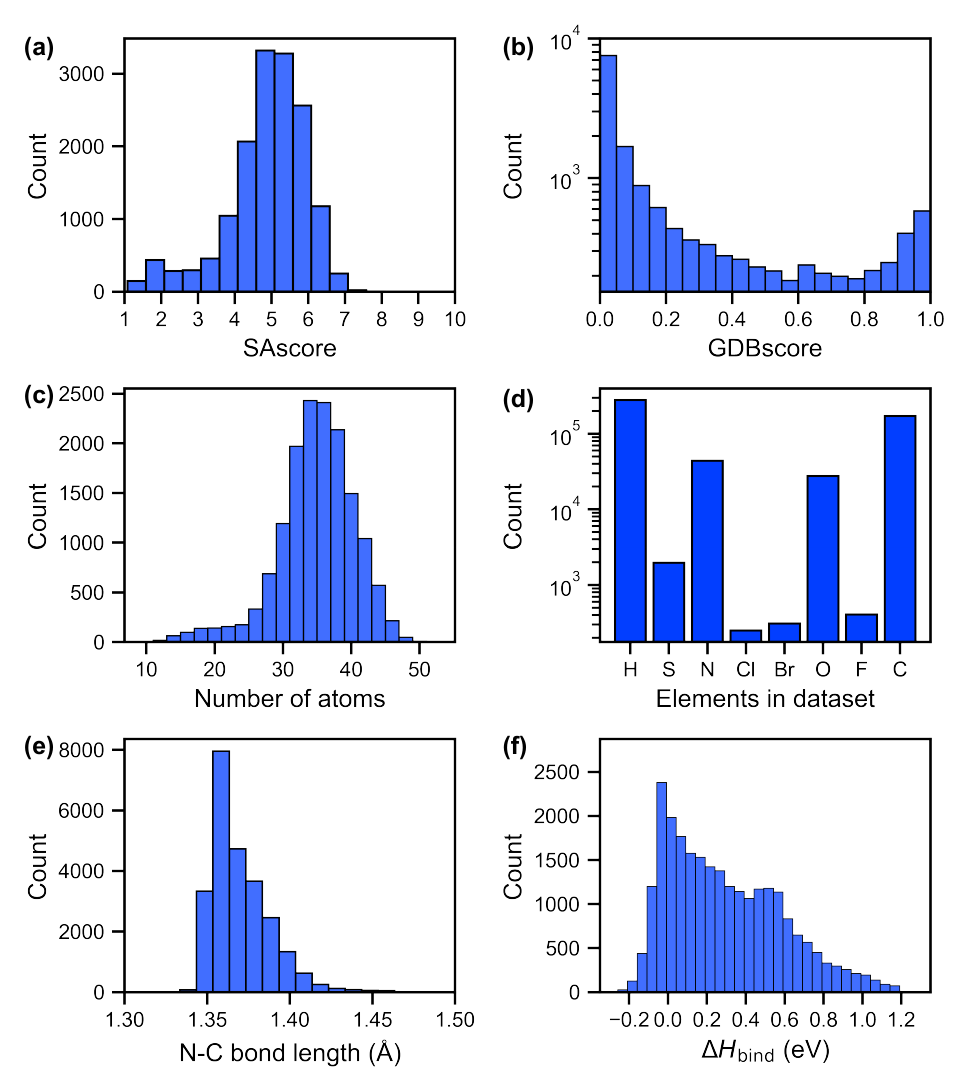}
	\caption[]{\label{fig:statistics} Histograms of selected statistics for DFT dataset used to train ML models:~(a) SAscore values, (b) GDBscore values, (c) Total number of atoms in each molecule, (d) Occurrence of each element in the dataset, (e) N-C bond lengths for the carbamic acid group in the ``child'' molecule in the DFT relaxed structure,~(f) DFT CO$_{2}$ binding enthalpies computed at B3LYP/6-311+G(d,p) level.}
\end{figure*}

\begin{figure}[ht]
	\centering
	\includegraphics[]{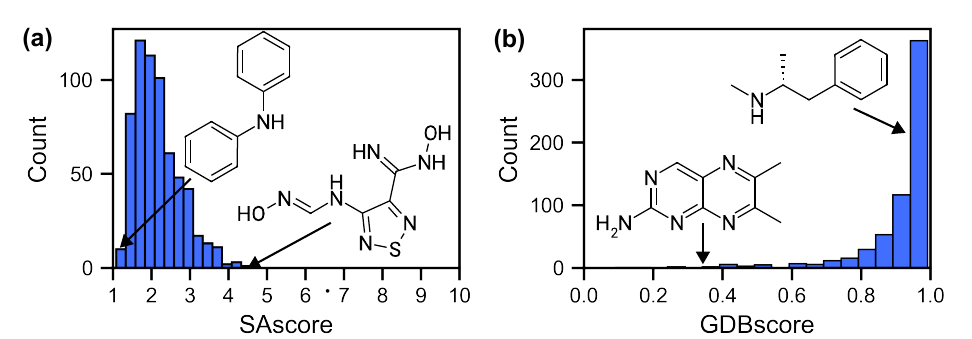}
	\caption{\label{fig:sa_and_ra_nist} Distributions of synthesizability scores for molecules with entries in the NIST database which are present in the combined DFT testing and training sets.~(a) SAscore distribution.~(b) GDBscore distribution. Insets show molecules selected from most unfavorable and favorable portions of the respective distributions.}
\end{figure}

Table~\ref{fig:statistics} shows selected statistics for our dataset.
The median number of atoms present in the GDB-17-DFT set is 35, although discounting hydrogen atoms the median number becomes 17.
The dominant elements present in the dataset are carbon, hydrogen, oxygen, nitrogen, and sulfur, with small numbers of fluorine, chlorine, and bromine present.
It should be noted that models trained on this dataset will be valid for similar regions of chemical space and are expected to become more and more error-prone as they are applied to molecules further from GDB-17-like molecules.
Also depicted in Figure~\ref{fig:statistics} is a histogram of N-C bond lengths, where N belongs to the primary or secondary amine and C belongs to the carbamate group which binds to the amine.
This shows a distribution centered tightly around a mean value of 1.37 \AA{} with a standard deviation of 0.02.

Regarding synthesizability metrics, the mean SAscore for the DFT dataset is 4.87 with a standard deviation of 1.08, while the mean GDBscore is 0.21 with a standard deviation of 0.30.
These suggest that most molecules from the DFT dataset are not synthetically accessible,\cite{Thakkar_2021,Ertl_2009} highlighting the utility of these synthesizability scores.
Furthermore, Figure~\ref{fig:sa_and_ra_nist} shows the distribution of SAscores and GDBscores for the subset of molecules in the DFT dataset which are also present in the NIST database.\cite{Huber}
Notably, most of these molecules have very favorable synthesizability scores for both metrics.
Over 95\% of the NIST subset has a GDBscore above 0.5, while no molecule in the NIST subset has an SAscore greater than 6, suggested to be a reasonable cutoff for estimating synthesizability with the SAscore~\cite{Ertl_2009}.
This validates the usage of these scores as synthesizability estimates, since they correctly estimate that experimentally studied molecules are synthesizable, or are at least natural products.

\clearpage

\subsection{BPEI Benchmark}

\begin{figure}[ht]
	\centering
	\includegraphics[width=\textwidth]{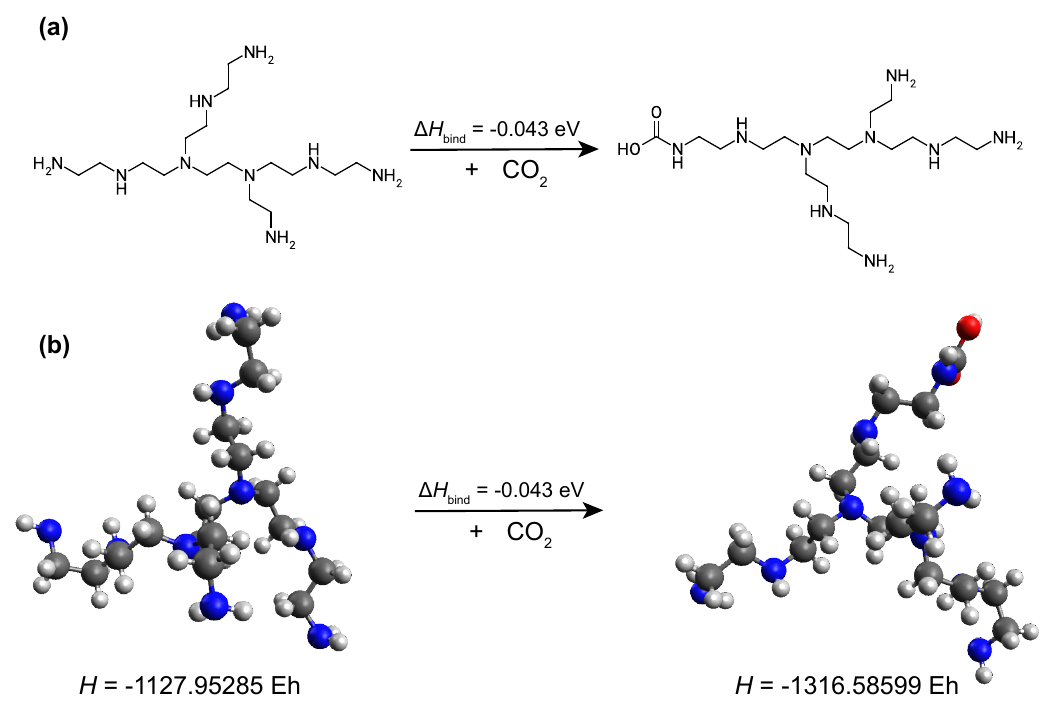}
	\caption{\label{fig:bpei_si} BPEI model used for CO$_{2}$ sorption benchmark. Structures are optimized and values computed using the DFT workflow outlined in main text. $\Delta H_{\mathrm{bind}}$ is given in eV, and enthalpy for each BPEI structure given in Hartrees. All values are computed at the B3LYP/6-311+G(d,p) level of theory. The optimized structures are available in .xyz format as additional files.}
\end{figure}

Figure~\ref{fig:bpei_si} shows 2D and 3D depictions of our BPEI reference model for the most thermodynamically favorable binding site.
Structures and energetics for the ``parent'' BPEI structure and all ``children'' are given as additional files.
For consistency in use as a theory-to-theory benchmark, these structures and energetics were computed using the exact same workflow discussed in the main text used to generate our DFT dataset.

\clearpage

\subsection{3D Model performance}

\begin{figure}[ht]
	\centering
	\includegraphics[]{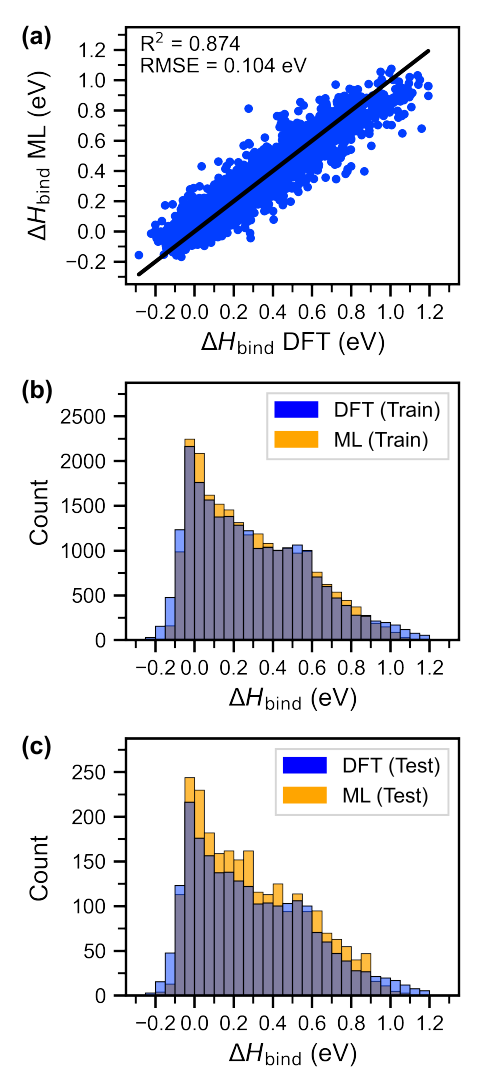}
	\caption[]{\label{fig:ml_results_3d} Performance of GBT model trained with 3D Mordred differential descriptors in addition to 2D differential descriptors.~(a) Truth plot for CO$_{2}$ binding enthalpies predicted by 3D model versus DFT enthalpies for the amine binding sites in the testing set.~(b) Histogram of CO$_{2}$ binding enthalpies for DFT and the 3D model for the amine binding sites in the training set.~(c) Histogram of binding enthalpies for DFT and ML for the testing set.}
\end{figure}

\begin{figure*}[ht]
	\centering
	\includegraphics[]{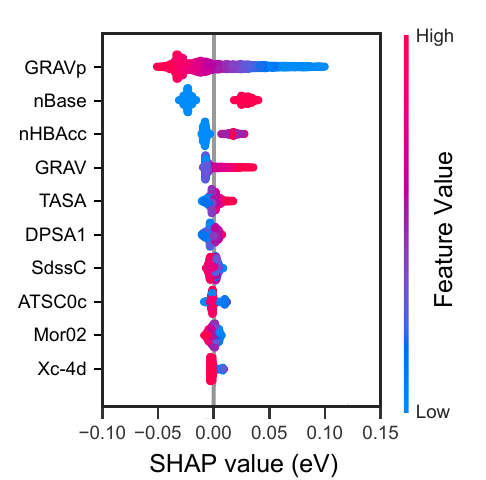}
	\caption{\label{fig:3DGBT_shap} Beeswarm plot of SHAP values for the top 10 most important descriptors for determining CO$_{2}$ binding enthalpies in amine-functionalized materials for the 3D GBT model.}
\end{figure*}

Figures~\ref{fig:ml_results_3d} shows the performance of a GBT model trained on both 2D and 3D Mordred descriptors and Figure~\ref{fig:3DGBT_shap} shows a beeswarm plot of the SHAP values for the top 10 descriptors of this model.
Differential descriptors were used for this model.
Rather than using the SMILES strings of the parent and child molecules to compute Mordred descriptors, the optimized DFT geometry of the parent and child are used to compute Mordred descriptors.
Thus, this model depends upon descriptors such as GRAVp which require a 3D geometry to be computed.

\clearpage

\subsection{Selected GDB-17-N Molecules}

\begin{figure*}[ht]
	\centering
	\includegraphics[]{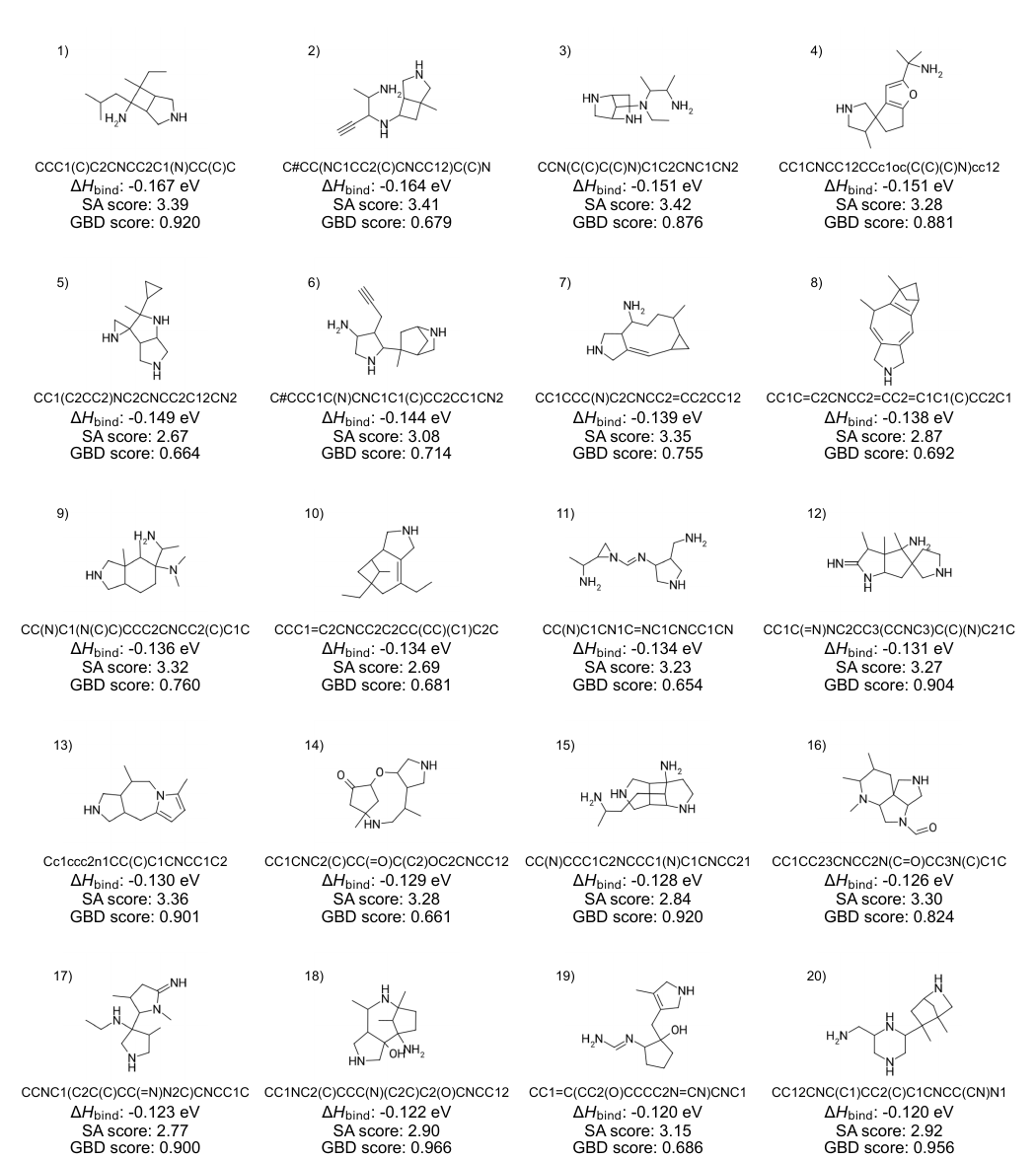}
	\caption{\label{fig:best_mols_matrix} The 20 molecules with most thermodynamically favorable $\Delta H_{\mathrm{bind}}$ values with favorable synthesizability scores in the GDB-17-amine set. GDBscores >0.64 and SAscores <3.4 are considered favorable, as discussed in main text.}
\end{figure*}

\begin{figure}[ht]
	\centering
	\includegraphics[]{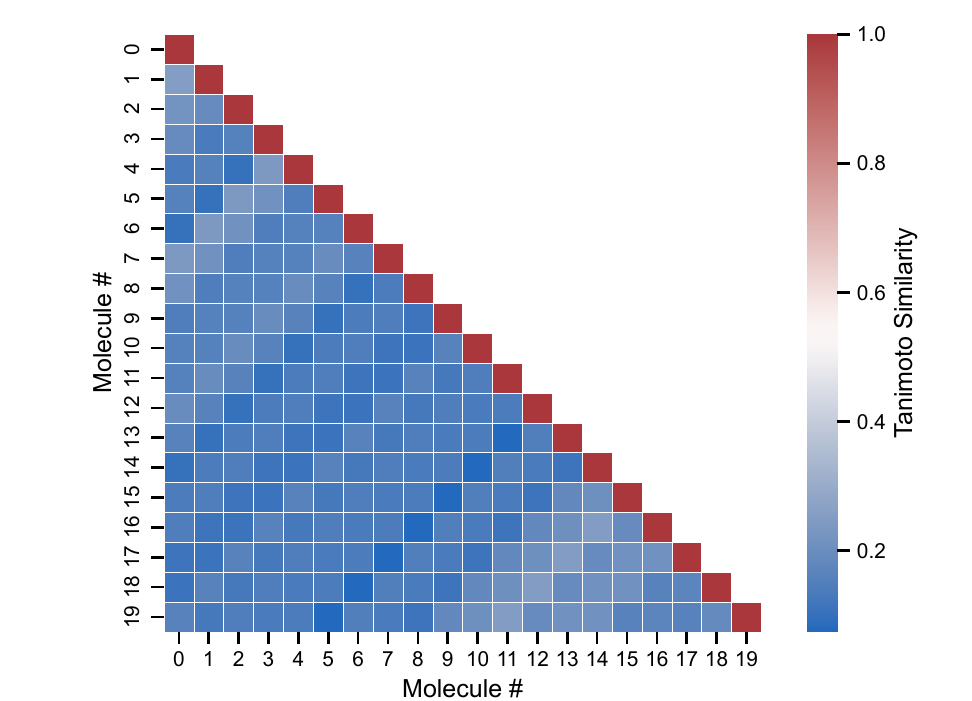}
	\caption{\label{fig:tanimoto_matrix} Tanimoto similarity matrix for the 20 molecules presented in Figure \ref{fig:best_mols_matrix}. The Tanimoto Similarity is computed using 2048 bit Morgan fingerprints derived from the SMILES of each molecule using a radius of 2.}
\end{figure}

Figure~\ref{fig:best_mols_matrix} gives twenty molecules with the most thermodynamically favorable CO$_{2}$ binding enthalpies and favorable synthesizability scores.
This table is provided as an illustrative example, and the full set of predicted binding enthalpies and synthesizability scores for the entire GDB-17-N set is given as an additional file.
Figure~\ref{fig:tanimoto_matrix} gives the Tanimoto similarity matrix for these molecules.

\clearpage

\subsection{Descriptor Correlations}

\begin{figure}[ht]
	\centering
	\includegraphics[]{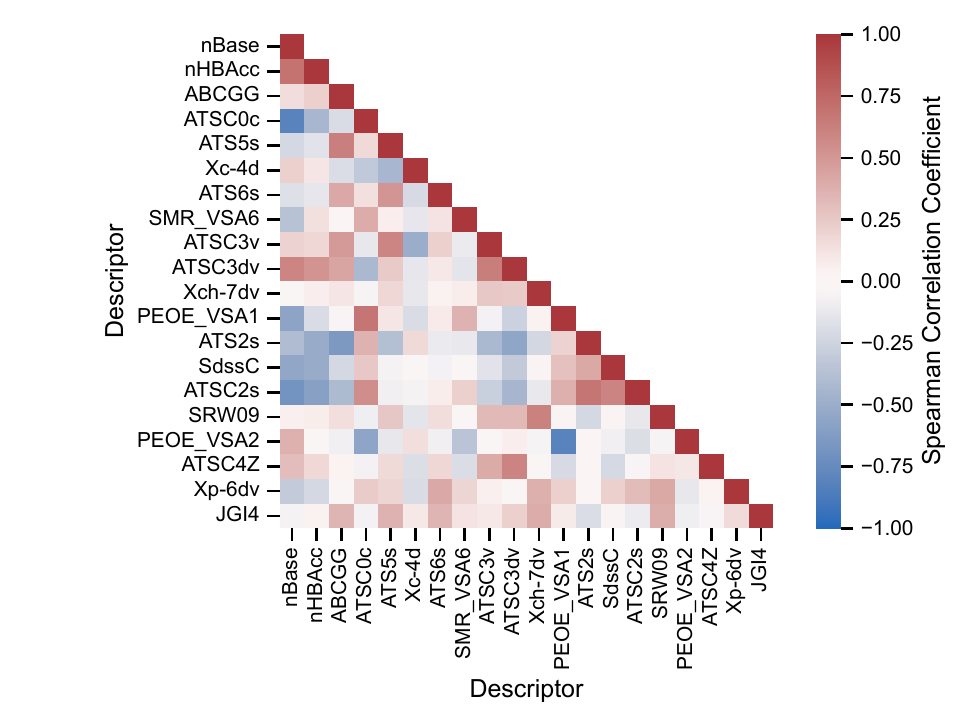}
	% \caption[]{\label{fig:descriptors_corr} Spearman correlation matrices for top 20 descriptors according to SHAP values for 2D Delta GBT.}
	\caption[]{\label{fig:descriptors_corr} Matrix of Spearman correlation coefficients for the best 2D GBT model. The top 20 most important descriptors according to SHAP value are given. A higher correlation coefficient indicates a closer correlation between the two descriptor values, while a negative coefficient indicates an inverse correlation and a value of zero indicates no correlation.}
\end{figure}

\begin{figure}[ht]
	\centering
	\includegraphics[]{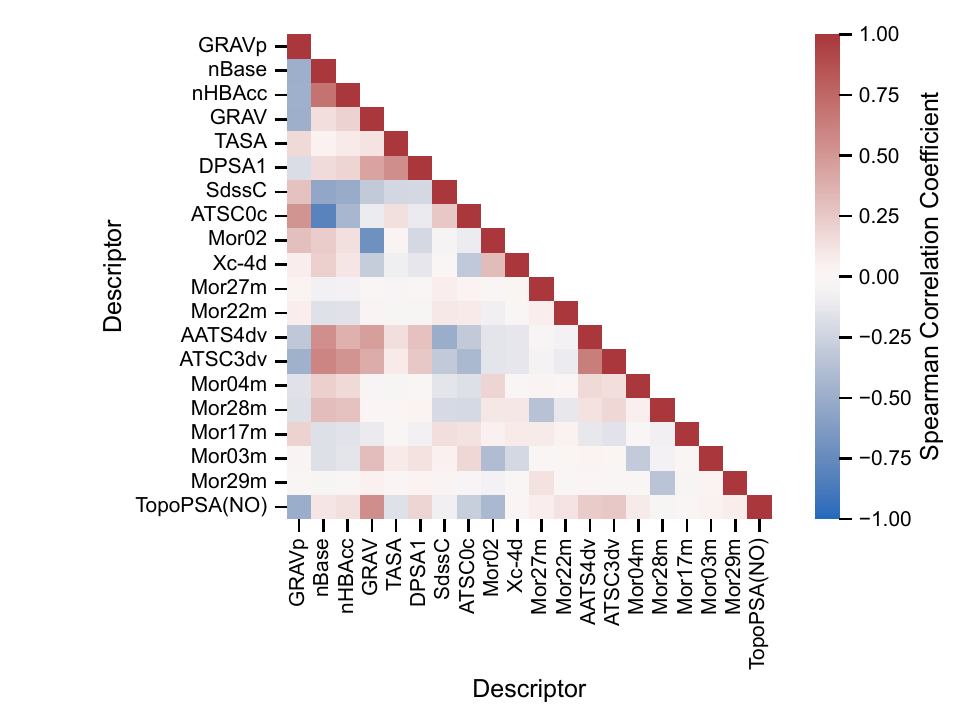}
	\caption[]{\label{fig:descriptors_corr_3D} Matrix of Spearman correlation coefficients for the 3D GBT model. The top 20 most important descriptors according to SHAP value are given.}
\end{figure}

Figures~\ref{fig:descriptors_corr}~and~\ref{fig:descriptors_corr_3D} give a matrix of Spearman correlation coefficients for our GBT models trained on 2D differential descriptors and 3D differential descriptors, respectively, for the top 20 most important descriptors according to SHAP value.

% \bibliography{/Users/megand/bib/references.bib}

% \pagebreak
\clearpage

\bibliography{references.bib}